# Designing a new spatial light modulator for holographic photostimulation

Janelle C Shane[1,*], Douglas J McKnight[1], Adrian Hill[2], Kevin Taberski[3], Steve Serati[1]

[1]Boulder Nonlinear Systems, Inc; [2]PlusPlus Software LLC; [3]Dynamic Engineering Corporation

**ABSTRACT**

Driven by the demands for speed and field of view in the holographic photostimulation community, we designed, built, and tested a liquid crystal on silicon (LCoS) spatial light modulator (SLM) with a 1536x1536 square pixel array and high-voltage LC drive. We discuss some of the engineering work that made the MacroSLM possible, including the custom FPGA board for handling huge data rates, the large pixel size for minimizing rolloff and crosstalk, and the temperature control to handle heating effects from the high-voltage controls and high-power laser illumination. We also designed an FPGA implementation of the overdrive method for increasing liquid crystal switching speed, allowing us to overcome the significant data bottlenecks that limit frame rates for large arrays. We demonstrate 500 Hz hologram-to-hologram speed at 1064 nm operating wavelength, and discuss the new science that these speeds and array sizes have enabled.

## 1. INTRODUCTION

Spatial light modulators (SLMs) have found widespread application due to their ability to modulate the amplitude and/or phase of light. One driving application in recent SLM development is photostimulation for optogenetics[1–6]. In photostimulation, engineered light-sensitive neurons are induced to fire when illuminated, allowing researchers to controllably activate neuronal circuits. Because neuronal circuits are interlaced with one another, selectively activating a particular network requires illuminating collections of individual cells rather than large areas. For this reason, many optogenetics researchers use liquid crystal on silicon (LCoS) SLMs to holographically convert a single excitation beam into multiple independently steerable beams.

The problem of holographic beamsteering for photostimulation is a particularly challenging one, since neuronal circuits can involve many hundreds of neurons, can cover a more than cubic millimeter in area[7], and have dynamics on the millisecond timescale. LCoS SLMs are the technology of choice for 3D holographic beamsteering, with high-resolution arrays of phase-modulating pixels that allow hundreds of beams to be independently steered and focused.

Most of the LCoS SLMs currently available are repurposed commercial display devices whose speeds are limited to around 60 Hz, much slower than the kilohertz timescales of neuronal circuits. A faster speed response is particularly important for closed-loop experiments, in which the displayed hologram is updated in real-time response to the dynamics of the neuronal circuit or other experimental feedback. This requires not only the liquid crystal (LC) pixels, but also the entire hologram data pipeline, to be designed to be capable of responding to triggers at these speeds. Additional design considerations such as pixel size, pixel count, and pixel crosstalk help determine the field of view that can be efficiently addressed with holographic beamsteering. Power handling is also an important consideration, since the near infrared (NIR) femtosecond sources used in optogenetics experiments can have average powers of multiple Watts, which can cause enough heating to affect the calibration of the LC response.

This paper describes the process of designing and testing a large-aperture high-speed SLM (called the MacroSLM) for applications in photostimulation (Figure 1). We consider the factors affecting 3D field of view, and present arguments for designing an SLM with large (20 micron) pixel size and high pixel count, resulting in a 1536 x 1536 array whose active area covers 30.7 x 30.7 mm. We also discuss the LC voltage response, and the use of high voltage (0-12V) analog addressing and transient high-voltage frames for maximizing LC response speed. We discuss backplane heating as a method for maintaining fast LC response and for controlling the effects of high-power laser illumination. We also describe a data handling strategy that allows us to make the most of the fast LC response speed, using a custom FPGA solution to calculate and load transient frames at 1250 Hz, and using an interruptible image download to allow flexible trigger response within 6 μs (+/-3 μs) of arbitrary trigger frequencies.

---

* jshane@bnonlinear.com; 303-604-0077; bnonlinear.com

The last part of this paper describes a method of measuring SLM frame rate. We characterize frame rate as the triggered rate at which the system can switch between typical spot-forming GS (Gerchberg-Saxton[8]) holograms while maintaining >90% of its steady-state diffraction efficiency. This provides a more complete picture than examining rise time or fall time alone, or examining the speed of a single phase transition, since it takes into account the entire pipeline of trigger response, data management, and pixel addressing as well as LC response over a realistic range of phase transitions. In addition, we make sure to measure speed in the NIR wavelength range where we plan to operate, since LC response is often more than 3x faster at visible wavelengths[9]. At 1064 nm, we easily achieve 500 Hz hologram-to-hologram frame rate.

## 2. DESIGN AND FABRICATION

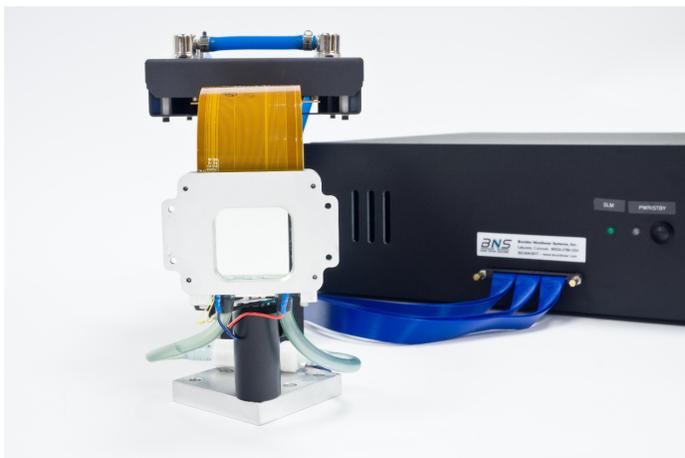

Figure 1. Photograph of the MacroSLM, along with its op-amp board and driver board.

**Pixel count**

For photostimulation, as well as for a large array of other holographic beamsteering applications, one major factor in the design of the system is 3D addressable field of view. Microscope magnification determines the extent of the field of view, but SLM design is important in determining the number of usefully addressable points, along with the excitation efficiency.

In general, the higher the SLM pixel count, the more individual points the SLM can address in a given microscope field of view. SLM pixel count also affects the diffraction efficiency of each steered beam. Looking at the simplest case of a single spot steered by a non-binary phase ramp, its 1st-order diffraction efficiency $\eta$ depends on Z, the number of phase pixels per ramp period, as follows[10]:

$$\eta_1 = \left[\frac{\sin\frac{\pi}{Z}}{\frac{\pi}{Z}}\right]^2$$

From this we can see that a spot steered to the theoretical max diffraction angle of the SLM array, corresponding to a pattern of period Z=2 pixels, can have 41% efficiency at most (Figure 2; Table 1). In practice, the efficiency is further reduced from this theoretical limit by the effects of dispersion, aberration, and pixel crosstalk.

Taking the pixel count effects into account, along with the effects of dispersion and aberration, the two-photon excitation efficiency of the steered beams are often decreased by 50% or more over the peripheral regions of the field of view[11].

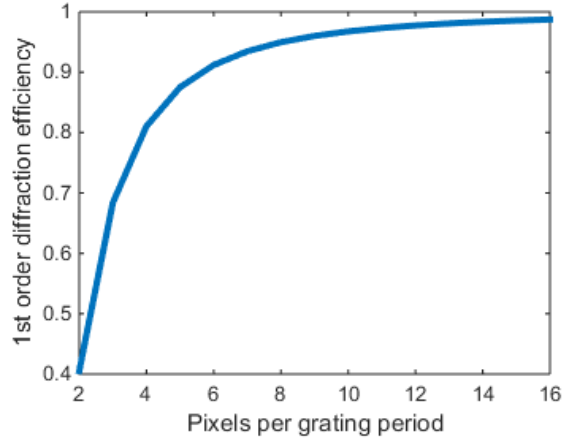

Figure 2. Theoretical 1st order diffraction efficiency of a blazed phase grating, given the number of phase levels (pixels) per grating period. Neglects the effects of pixel crosstalk, error in reproducing phase levels, and imaging system nonidealities including chromatic dispersion.

Table 1. Theoretical 1st order diffraction efficiency of a blazed phase grating, given the number of phase levels (pixels) per grating period. Neglects the effects of pixel crosstalk, error in reproducing phase levels, and imaging system nonidealities including chromatic dispersion.

| Pixels per grating period | 1st order diffraction efficiency | Pixels per grating period | 1st order diffraction efficiency |
|---|---|---|---|
| 2 | 0.4053 | 10 | 0.9675 |
| 3 | 0.6839 | 11 | 0.9731 |
| 4 | 0.8106 | 12 | 0.9774 |
| 5 | 0.8751 | 13 | 0.9807 |
| 6 | 0.9119 | 14 | 0.9833 |
| 7 | 0.9346 | 15 | 0.9855 |
| 8 | 0.9496 | 16 | 0.9872 |
| 9 | 0.9600 | 17 | 0.9887 |

In a similar manner, the number of available SLM pixels affects the SLM's ability to efficiently add focus/defocus to the steered spots, which affects the addressable field of view in the axial direction. The required fringe spacing for a quadratic phase function decreases very quickly as the amount of focus/defocus increases. For experiments that require stimulating multiple layers in the brain, some of which may be separated by many hundreds of microns, the demand on SLM pixel count is particularly high[11]. Some researchers also take advantage of an SLM's ability to apply aberration-correction phase modulation; the range and accuracy here is also affected by the SLM's pixel count.

In practice, the pixel count of SLMs is limited by two major factors: 1) fabrication cost and 2) data handling. As the pixel count increases, the cost of fabricating the large silicon backplane increases, and yield-related losses can potentially increase. Larger arrays may exceed the single-shot write area of the fabrication foundry, so care must be taken to avoid misalignments and other nonuniformities when stitching together write areas. Data handling is also a limitation, since large arrays of pixel data must be transferred from the computer to the SLM and the individual voltages loaded onto each SLM pixel. Any associated voltage calibration and transient frame calculations are also considerably slowed by large arrays. The hardware and development costs associated with performing these operations at hundreds or thousands of Hz are considerable.

For our design, we chose a square array size of 1536x1536 pixels, as a compromise between these factors. Although most LCoS SLM arrays available are rectangular, this is a reflection of their being repurposed display devices rather than any advantage for holographic beamsteering. In practice, SLMs in holographic beamsteering are usually illuminated by circular beams, so the most relevant pixel count is that of the shortest dimension. A square array takes maximum advantage of available backplane space and avoids slowing frame rates by loading voltages onto unused pixels.

**Pixel size**

Although a larger pixel pitch corresponds to smaller diffraction angles, the field of view is not affected as long as the magnification of the optical system is designed to image the SLM to the size of the pupil of the objective lens. For the MacroSLM we choose a relatively large 20 μm pixel pitch. There are several reasons to prefer a large pixel pitch for holographic beamsteering.

**Fill factor:** Although in the previous section we focused on the +1 diffraction order, any pixelated SLM diffracts to multiple orders in the optical far-field. In most applications it is preferred for as much power as possible to be directed to the +1 order. If we model the SLM pixel as a perfectly flat square mirror with a non-reflective gap between it and its neighbors, then it can be shown that this type of diffraction efficiency is proportional to the square of the area fill factor (the proportion of the device surface that is occupied by mirror). Assuming a constant inter-mirror gap, larger pixels result in a larger fill factor. For our 20 μm pixel pitch device, the inter-mirror gap is just under 1 μm, yielding a fill factor of a little over 95%.

**Pixel crosstalk:** An additional motivation for using large pixels is the hologram-dependent distortion of the LC alignment at the pixel edges. These distortions are known as fringing fields or crosstalk and acts as a low-pass filter on the displayed hologram. Crosstalk can therefore have a dominant effect on the SLM's diffraction efficiency at higher steering angles or when steering multiple spots. A device with 10 μm pixel pitch and an 0.5 μm inter-mirror gap would have the same fill factor as a device with 20 μm pixel pitch and a 1 μm inter-mirror gap, but might have reduced diffraction efficiency due to the greater contribution of crosstalk at the pixel edges.

Crosstalk's effects are most evident in diminishing the diffraction efficiency toward the edge of the field of view. There are techniques for compensating for this, including accounting for crosstalk in the part of the Gerchberg-Saxton (GS) algorithm that predicts the diffracted spots[12]. However, this can impair the GS algorithm's convergence, leading to calculation times that are more than 4x longer in many cases, and the effects of crosstalk are diminished rather than eliminated.

It is important to note that crosstalk effects are very small if a uniform phase value is written to all the pixels, so a small-pixel SLM may exhibit acceptable, or good, diffraction efficiency under simple test conditions but perform less well with high spatial-frequency hologram data where crosstalk dominates.

**Stability:** A larger pixel pitch also allows for larger pixel storage capacitance. Due to finite pixel capacitance, the SLM pixels do not perfectly maintain their loaded voltages, but experience a gradual voltage decay. LCoS SLMs are therefore designed to periodically refresh the pixel voltages. Between refreshes, the sagging pixel voltage results in a periodic change in pixel phase, which in turn produces an intensity ripple in the diffracted spots. A larger pixel storage capacitance will reduce the amount of intensity ripple, producing a more stable illumination. The degree of intensity ripple for a given displayed phase value also depends on whether that phase value is in a steep region of the nonlinear phase-voltage response curve; if the SLM is displaying a phase value in a slowly-changing region of the response curve, the phase ripple will be minimal even at slow refresh rates.

A faster refresh rate also reduces the amount of intensity ripple. LCoS SLMs that are repurposed from display applications tend to have low refresh rates of around 60 Hz, while SLMs designed for low-ripple phase modulation can have refresh rates of 1 kHz or higher. For the MacroSLM we chose a refresh rate of 1250 Hz, limited by the engineering effort involved in loading voltages onto a large 1536x1536 pixel array.

**Large-aperture array:** The combination of a large pixel pitch and high pixel count leads to large device sizes, in our case an active array area of 30.7 x 30.7 mm. Although this necessitates larger optics in the system, some photostimulation setups are already moving to low-magnification high-NA imaging that require microscope objectives with back apertures of 30 mm or more[13–15]. Because a SLM is typically imaged onto a microscope objective's back aperture, a closer match between SLM and microscope objective aperture means less (or even zero) magnification change, and fewer of the aberrations associated with magnification change.

Power handling is also an advantage of large-aperture SLM arrays, because the power incident on the SLM is spread over a larger array area. This increases the total power-handling capacity of the device, and also lowers the heating effects of incident illumination.

**Voltage range**

In LCoS SLMs, a voltage is applied across an LC-containing cell using two electrodes. One electrode is the pixelated silicon backplane, while the other is a transparent electrode on the SLM's coverglass. The applied voltage causes the birefringent LC molecules to rotate, producing voltage-dependent phase retardation in the light that passes through the cell. The speed of the LC rotation determines the switching speed of the SLM. The transition from an initial phase $\phi_0$ to a final phase $\phi_1$ beginning at time $t_0$ can be approximated by an exponential phase

$$\phi(t) = \phi_0 + (\phi_1 - \phi_0)\left(1 - e^{-\frac{t-t_0}{\tau}}\right)$$

where the time constant $\tau$ depends on the direction of the phase transition and on LC viscosity[16]. The larger the difference between initial and final phase, the faster the phase change initially occurs, slowing as it nears its steady-state value. Therefore, one way to increase switching speed is to transiently apply large voltages until the phase reaches its target value, a technique known as overdrive. The higher the voltage the SLM can apply, the higher the speed boost from overdrive[16].

The use of overdrive requires an SLM with a kHz or higher refresh rate, since the SLM will need to display multiple images during each phase transition, swapping overdrive voltages for target voltages as each pixel reaches its target phase. The SLM also needs a larger voltage range than that required to reach a full wave of phase modulation.

As operating voltage increases, the amplifying electronics become more complex, and self-heating effects can cause enough temperature rise for the LC to become isotropic and stop switching, unless temperature is controlled.

For our SLM we choose high-voltage (0 – 12 V analog) pixel addressing for fast LC response and strong overdrive. We optimized the SLM thickness to about 1.5 waves of phase stroke at our NIR (~1064 nm) target wavelength. We used the entire 0.5 wave excess phase stroke for overdrive speed boost in the slower (relaxation) direction, where full-range phase transitions are on the order of milliseconds. In the fast direction, phase transitions are mostly under 1 ms, even without overdrive speed boost. We also included phase wrapping, in which at each pixel we evaluate the speed at which it can reach its destination or an equivalent 1 wave multiple of that destination, and choose the fastest transition available.

**Backplane heating**

Another important factor of our design is the ability to maintain the LC temperature at 45 C with the use of backplane Peltier heating/cooling. This temperature is high enough that LC viscosity is low, yet not high enough that our available phase stroke is significantly reduced. Active heating/cooling also allows us to adjust for the varying heating effects of high-power laser illumination, as well as for the effects of self-heating. The timing of our transient overdrive voltages is very temperature-dependent, as is the phase stroke to a lesser extent. Maintaining a constant LC temperature is important for maintaining the calibration of our overdrive timings, as well as our voltage-phase relationship.

**Data pipeline**

Data handling is another significant part of our speed effort, since the system must be capable of calculating the required transient voltages to achieve fast LC switching from phase to phase at each pixel, while loading the transient 1536x1536 images onto the SLM pixels at ~1250 Hz continuous frame rate. We use a custom FPGA solution for handling these high data rates, including on-board storage of 2045 images, on-board application of spatially-varying voltage calibrations, and on-board calculation of individual transient voltages for every pixel. The transient voltages must be calculated on-the-fly to allow for flexible switching between stored target holograms; otherwise, the stored images could only be displayed in a single predetermined order.

The driver board receives data over PCIe on cable to a Xilinx Kintex-7 primary FPGA. This FPGA distributes the data to 8 secondary Kintex-7 FPGAs using the Xilinx Aurora high-speed serial interface. Each secondary FPGA performs the overdrive processing for, and supplies the data to, its own section of the SLM. The primary FPGA also contains a Microblaze soft microcontroller that performs a number of additional functions, such as loading certain parameters over $I^2C$, temperature monitoring, and automatic safety-shutdown for both the driver board and SLM head. Software control is provided with an SDK with a C-language interface that can be called from LabView, Matlab, Python, or C/C++. We have also developed a GUI interface for interactive control of the SLM.

With this data-handling pipeline, the ultimate limitation on our hologram display frame rate is the response time of the LC, rather than the response of the electronics. Our system can be triggered at arbitrary rates exceeding 1 kHz.

**Flexible triggering**

Another requirement for photostimulation (and other applications) is the ability for the system to respond flexibly to triggers at arbitrary intervals, rather than at a fixed interval. For closed-loop experiments in particular, the system should be able to switch illumination patterns in response to experimental phenomena/animal behavior, even if the interval varies from a millisecond to tens of seconds within the same sequence of images.

A data-handling system that enables interruptible image downloads is a key requirement for this. Otherwise, if triggers can only be handled whenever an image download is complete, this restricts the trigger interval to integer multiples of the SLM's base refresh rate. Latency and jitter can both be significant, especially for SLMs with low refresh rates. An SLM with a 60 Hz refresh rate could have a trigger response latency that varies from 0 to 17 ms depending on when the trigger arrives within the refresh cycle.

With the ability to interrupt the download of an image onto the SLM's pixels, the latency between a trigger arriving and the voltage changing on our SLM is 6 μs with a range of +/-3 μs, so that the transition to a new hologram can be very predictably initiated. This latency is well within the LC response time and is therefore insignificant under photostimulation experimental conditions. Figure 3 demonstrates the system's response to a trigger signal whose frequency sweeps from 500 Hz to 100 Hz, with the LC switching between spot-forming holograms in response to each falling edge (more details on this setup in Section 3).

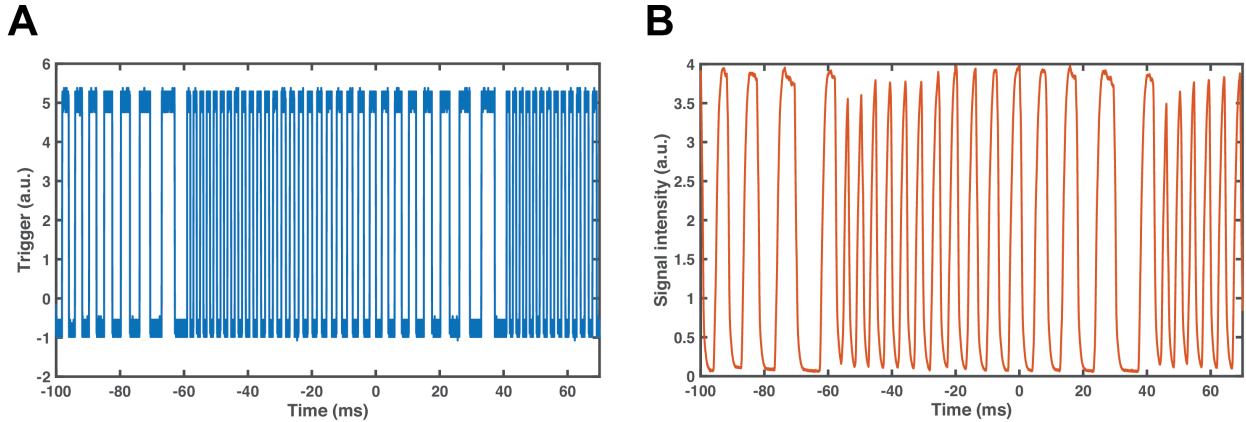

Figure 3. Triggering hologram transitions at arbitrary intervals. A) Trigger signal (SLM triggers on falling edge) with frequency sweeping from 500 Hz to 100 Hz at a 10 Hz period. B) SLM optical response. Test holograms are 8-spot Gerchberg-Saxton (GS) spot-steering holograms applied to the SLM 1536 x 1536 array. One hologram steers a spot to a single target location where we place our photodetector (Thorlabs InGaAs PDA50B), plus 7 other random locations. Another set of 10 holograms steers spots to 8 random locations with the photodetector's location excluded.

## 3. SPEED CHARACTERIZATION

We characterize the speed of this SLM as the triggered rate at which the system can switch between typical spot-forming GS (Gerchberg-Saxton[8]) holograms while maintaining >90% of its steady-state efficiency. This provides a more complete and relevant picture than examining rise time or fall time alone, or examining the speed of a single phase transition, since there are significant differences between the speeds of individual phase transitions, and also between spot formation versus destruction. Instead, it takes into account the entire pipeline of trigger response, data management, and pixel addressing as well as LC response over a range of phase transitions. In addition, we make sure to measure speed at 1064nm, in the wavelength range where we plan to operate, since LC response is often more than 3x faster at visible wavelengths[9].

Figure 4 shows the measured switching between a series of 11 8-spot GS holograms, one set of which places a spot at our photodetector location plus 7 other random locations, and the other set of which avoids placing a spot at the photodetector. We trigger the SLM controller at 100 Hz for a slow-speed reference, and then at our test frame rate. At 500 Hz our peak-to-peak signal amplitude is 96% of the slow-speed amplitude, well above the 90% target we use to define our hologram-to-hologram frame rate. (At 600 Hz, our peak-to-peak amplitude is 88%, just below this target threshold.) Note that we choose these trigger frequencies for convenience; since our image download is interruptible, we're not restricted to integer multiples of a base refresh rate.

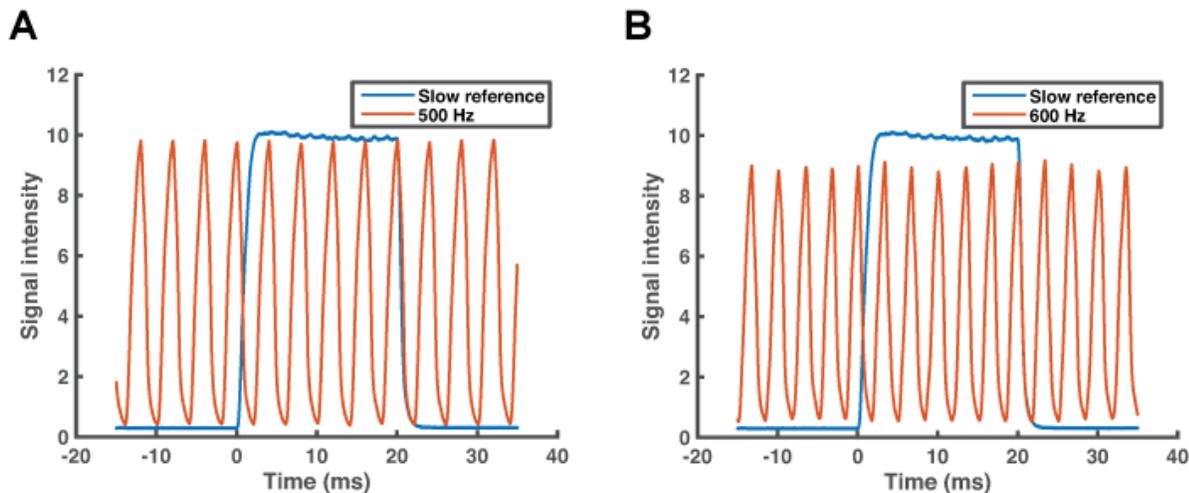

Figure 4. Hologram-to-hologram switching of the 1536 x 1536 MacroSLM. Operating wavelength is 1064 nm (LC response at visible wavelengths is typically >3x faster). Test holograms are 8-spot Gerchberg-Saxton (GS) spot-steering holograms applied to the SLM array. One hologram steers a spot to a single target location where we place our photodetector (Thorlabs InGaAs PDA50B), plus 7 other random locations. Another set of 10 holograms steers spots to 8 random locations with the photodetector's location excluded. In this way, we sample a representative variety of phase transitions and complex holograms. Transient voltage ("overdrive") frames are calculated on-the-fly by on-board FPGAs and delivered at ~1250 Hz. Blue: System triggered at 100 Hz, indicating the diffracted signal strength at moderate speeds. A) Red: System triggered at 500 Hz. Peak-to-peak signal amplitude is 96% of the slow-speed amplitude. B) Red: System triggered at 600 Hz. Peak-to-peak signal amplitude is 88% of the slow-speed amplitude.

## 4. CONCLUSION

Here we discuss the process of designing an LCoS SLM specifically for photostimulation applications, where requirements for speed, field of view, and trigger responsiveness are driving the development of advanced SLMs. We designed a 1536 x 1536 pixel SLM that can be flexibly triggered at arbitrary intervals of < 1 ms, with a latency of 6 μs +/- 3 μs. Achieving these frame rates required a significant data handling effort, particularly with on-board calculation of transient overdrive frames.

At the NIR wavelengths usually required for photostimulation, LC responsiveness is what ultimately limits the maximum frame rate at which the SLM can display holographic spot patterns. At 1064 nm, our SLM displays GS holograms at 500 Hz while maintaining 96% of its slow-speed diffraction efficiency. At 600 Hz triggering, it maintains 88% of its slow-speed amplitude. For these measurements we took care to use GS holograms rather than simple grating patterns, since simpler patterns don't sample the full range of spatial frequencies and phase transitions required for a multispot photostimulation experiment. At visible wavelengths, we would expect LC response to be more than 3x faster than the speeds we report at 1064 nm.

This system represents an advance in holographic photostimulation speed and field of view and has been recently used to achieve frame rates of >1 kHz through the MultiSLM method of multiplexing two triggered systems[17]. We expect that as other labs begin using the MacroSLM, it will enable dynamic experiments that push the state of the art in multiple domains.

Future work will improve this performance further. New materials may push the LC response speed higher, producing a faster response at current operating temperatures or allowing operation at more-elevated temperatures where LC viscosity is lower. An array with lower pixel count can allow a faster refresh rate and therefore finer control of overdrive, which may improve speed and reduce phase ripple at the expense of fewer addressable spatial points. As the demands of photostimulation push progress in LCoS SLM development, it will open up new applications for LCoS devices.